\documentclass[sigconf,screen,nonacm]{acmart}

\AtBeginDocument{%
  \providecommand\BibTeX{{%
    \normalfont B\kern-0.5em{\scshape i\kern-0.25em b}\kern-0.8em\TeX}}}

\setcopyright{rightsretained}

\acmConference[arXiv Preprint]{arXiv Preprint}{January 2026}{Adelaide, Australia}
\acmYear{2026}
\acmDOI{} 
\acmISBN{} 

\usepackage{xcolor}
\usepackage{todonotes}
\usepackage{graphicx}
\usepackage{csquotes}
\usepackage{adjustbox}
\usepackage{soul}
\usepackage{colortbl}
\usepackage[table]{xcolor}
\definecolor{Gray}{gray}{0.9}

\begin{document}

\title{Who Fails Where? LLM and Human Error Patterns in Endometriosis Ultrasound Report Extraction}




\author{Haiyi Li}
\email{a1949007@adelaide.edu.au}
\affiliation{%
  \institution{Univ. of Adelaide}
  \city{Adelaide}
  \country{Australia}
}

\author{Yutong Li}
\email{a1948101@adelaide.edu.au}
\affiliation{%
  \institution{Univ. of Adelaide}
  \city{Adelaide}
  \country{Australia}
}

\author{Yiheng Chi}
\email{yiheng.chi@student.adelaide.edu.au}
\affiliation{%
  \institution{Univ. of Adelaide}
  \city{Adelaide}
  \country{Australia}
}


\author{Alison Deslandes}
\email{alison.deslandes@adelaide.edu.au}
\affiliation{%
  \institution{Robinson Inst., Univ. of Adelaide}
  \city{Adelaide}
  \country{Australia}
}

\author{Mathew Leonardi}
\email{leonam@mcmaster.ca}
\affiliation{%
  \institution{McMaster University}
  \city{Hamilton}
  \country{Canada}
}

\author{Shay Freger}
\email{Fregers@mcmaster.ca}
\affiliation{%
  \institution{McMaster University}
  \city{Hamilton}
  \country{Canada}
}

\author{Yuan Zhang}
\email{yuan.zhang01@adelaide.edu.au}
\affiliation{%
  \institution{Robinson Inst., Univ. of Adelaide}
  \city{Adelaide}
  \country{Australia}
}


\author{Jodie Avery}
\email{jodie.avery@adelaide.edu.au}
\affiliation{%
  \institution{Robinson Inst., Univ. of Adelaide}
  \city{Adelaide}
  \country{Australia}
}

\author{Mary Louise Hull}
\email{louise.hull@adelaide.edu.au}
\affiliation{%
  \institution{Robinson Inst., Univ. of Adelaide}
  \city{Adelaide}
  \country{Australia}
}

\author{Hsiang-Ting Chen}
\email{tim.chen@adelaide.edu.au}
\affiliation{%
  \institution{Univ. of Adelaide}
  \city{Adelaide}
  \state{SA}
  \country{Australia}
}

\renewcommand{\shortauthors}{Li et al.}

\begin{abstract}
In this study, we evaluate locally deployed large language models (LLMs) for converting unstructured endometriosis transvaginal ultrasound (eTVUS) reports into structured data. Across 49 de-identified reports, we compared three on-premise LLMs (7B/8B and 20B parameters) against expert human extraction using a 185-field schema. The 20B model achieved the highest mean accuracy (86.02\%), substantially outperforming the smaller models. Crucially, LLMs and humans exhibited complementary error patterns: the LLM excelled on structured fields (date formatting, measurement decomposition) where humans made protocol errors, while humans demonstrated superior performance on interpretive fields involving negation and clinical terminology. Targeted prompt engineering yielded only marginal gains, indicating that these errors reflect model limitations rather than instruction gaps. These findings support a human-in-the-loop workflow in which the LLM generates structured drafts, automated validation flags rule-verifiable errors, and human review focuses on fields requiring clinical interpretation.
\end{abstract}



\maketitle


\section{Introduction}
Free-text ultrasound reports contain clinically valuable information, but key variables are embedded in heterogeneous narrative styles and local formatting conventions, limiting their use in analytics, model training, and auditing \cite{Wang2018,Langlotz2006,info:doi/10.2196/12239}. Across clinical domains, this structural barrier complicates secondary use and necessitates substantial manual abstraction \cite{Wang2018,castro2017automated,davis2013automated,savova2017deepphe}. 
In settings where privacy requirements preclude cloud-based processing, extraction remains a manual, safety-critical task: abstractors must interpret clinical content while enforcing protocol constraints such as field decomposition, formatting standards, and missingness conventions \cite{Langlotz2006,Holste2023}. 
Our contextual inquiry identified recurring risk points, including terminology variation, inconsistent report detail, and verification-heavy routines that induce fatigue and increase silent transcription and field-alignment errors \cite{Langlotz2006,Holste2023}. 
These observations suggest that effective support tools should prioritise reviewability and accountability over full automation, enabling practitioners to calibrate their reliance on algorithmic assistance within real workflows \cite{fitzpatrick2013review,cai2019hello,beede2020human}.

Locally deployable LLMs offer a practical means of scaling abstraction without transmitting sensitive reports to external services \cite{Brown2020,Li2023}. Recent studies demonstrate that LLMs can perform few-shot clinical extraction with substantial medical knowledge \cite{agrawal2022large,Singhal2023}, yet they also produce well-formed outputs that are semantically incorrect, failures that are difficult to detect from the output alone \cite{Agrawal2023,Holste2023}. In structured reporting, this problem is acute: schema-compliant responses may still mishandle negation, map terms to incorrect categories, or misinterpret context-dependent findings. 
Without visible indicators of error, users struggle to calibrate trust, leading to both over-reliance and under-reliance on model outputs \cite{bucinca2021trust,bussone2015role}. 
Research on clinical AI deployments reinforces this concern, showing that operational success depends less on standalone accuracy than on workflow integration and mechanisms that direct reviewer attention to high-risk outputs \cite{beede2020human,cai2019hello}. Designing such mechanisms requires understanding where LLMs and humans each fail.

This study addresses that gap by investigating the error patterns of local LLMs and human abstractors to inform design guidelines for human-AI collaborative abstraction systems. Using 49 de-identified eTVUS reports and a 185-field extraction schema, we benchmark three on-premise models (7B to 20B parameters) against expert-verified human abstraction. Our analysis targets two dimensions with direct design implications: report-level variance, which governs review effort and suggests where batch processing is viable, and field-type error patterns, which reveal where human judgment remains essential and should be preserved. We find that LLMs and humans fail in complementary ways, motivating a division of labour in which automated validation catches mechanical errors and risk-based triage routes semantically sensitive fields to human review. We also test whether targeted prompting can reduce errors on critical fields; the limited and inconsistent gains suggest that workflow-level safeguards, rather than prompt refinement, offer the more reliable mechanism for managing extraction risk.

\section{Experiment} 
\label{sec:Method}
\subsection{Data and Schema} 
The dataset consisted of unstructured, de-identified sonologists reports obtained from a specialized gynecology and obstetrics ultrasound clinic in Canada. These reports were heterogeneous, containing both structured data fields and free-text ultrasound narratives. To prepare the data for the pipeline, each report, originally in PDF format, was converted to plain text. We used a layout-preserving extraction process designed to retain semantic content while removing extraneous metadata and formatting artifacts. 

The extraction target was defined by a structured endometriosis centric schema. This schema was created by programmatically transforming the header row of a reference Excel data dictionary into a concise JSON schema. This file defined all key fields, their data types, and output format constraints, serving as the ground truth structure for both model prompting and final evaluation \cite{agrawal2022large,savova2017deepphe}. 
The reference Excel file contained 185 fields in total. Each field was programmatically assigned a data type based on its values and intended use. The schema included five major data types: Numeric (6 fields), Date (2 fields), Text (19 fields), and Categorical (157 fields). The majority of fields were categorical, typically representing controlled vocabularies or discrete clinical options, while a smaller subset are free-text or numeric entries. This distribution reflected the highly structured nature of the target schema and the clinical emphasis on standardized reporting.

\subsection{Extraction Pipeline} 
We designed an on-premise extraction pipeline to ensure full patient privacy and data sovereignty. The entire system operated offline, without reliance on external APIs, and is deployed on commodity hardware. 
The pipeline was built on the OLLAMA platform, using three different LLM models \textbf{gpt-oss:20b}, \textbf{llama3-8b} and \textbf{mistral-7b}. 
The workflow proceeded as follows: 
\begin{enumerate} 
    \item \textbf{Schema-Guided Prompting:} Each plain-text report was processed sequentially in batch mode. For every report, the model received the full textual content along with the JSON schema embedded within the prompt as an instructional template. 
    \item \textbf{Inference:} The LLM generated a structured JSON object containing the extracted field-specific values. This output was saved as an intermediate file for validation. 
    \item \textbf{Validation and Post-processing:} A rule-based validation layer was applied to the JSON outputs. This step normalized missing value indicators (e.g., \texttt{0}, \texttt{NA}, and empty strings), harmonized categorical variables to a controlled vocabulary, and verified field completeness and data-type conformity \cite{leyh2018deep}. 
\end{enumerate} 
The experiment ran on a personal computer equipped with an NVIDIA RTX 3090 GPU (24GB VRAM).

\begin{figure}[htbp]
  \centering
  \includegraphics[width=\columnwidth]{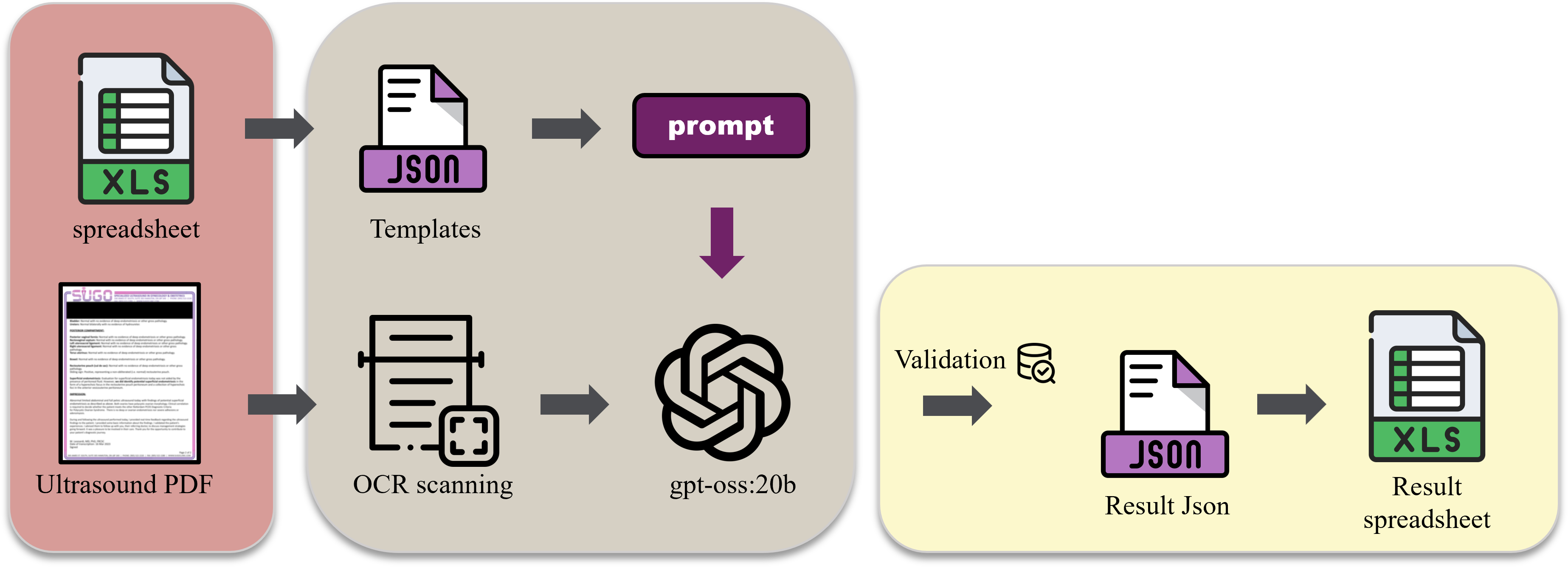}
  \caption{Overview of the on-premise workflow. De-identified ultrasound PDFs and the organization’s spreadsheet template are processed locally by a layout-aware LLM backend to produce a schema-aligned structured JSON draft. A field stratification strategy and rule-based validation prioritize semantically sensitive fields for review. An interactive human-in-the-loop interface supports rapid trace-back from each field to relevant PDF text anchors, enabling verification and correction before exporting a verified structured dataset for clinical research use.}
  \Description{Three-panel workflow diagram. Left: automated backend pipeline taking batch ultrasound PDFs and a spreadsheet template, running LLM processing and a field classification strategy to produce structured JSON. Middle: an interactive review interface showing a PDF viewer and a structured form with a callout indicating one-click trace-back to PDF text anchors. Right: final verified structured database/CSV output.}
  \label{fig:workflow_hitl}
  \vspace{-4mm}
\end{figure}

\section{Preliminary Evaluation}
\label{sec:Result}

\textbf{Scoring and accuracy definition.}
We report \emph{field-level accuracy} scored against an expert-verified reference (\emph{Verified Truth}). For each report, we score each of the 185 schema fields as correct (1) if the normalized prediction matches the normalized reference value, and incorrect (0) otherwise; report-level accuracy is the mean across fields, and we report the mean and standard deviation (SD) across 49 reports.
For \textbf{numeric/date fields}, we convert outputs into a canonical representation and compare for equality in that canonical form. For \textbf{text/categorical fields}, we apply lightweight normalization (case-folding plus whitespace and punctuation normalization) and then require an exact match.
Missing information is represented by an explicit token \texttt{NOT\_MENTION}; a field is counted as correct when both prediction and reference indicate \texttt{NOT\_MENTION}, and incorrect when one indicates missingness while the other provides a value. For protocol fields where \texttt{0} and \texttt{NA} both denote \emph{not detected / not recorded}, we treat \texttt{0} and \texttt{NA} as the same missingness state during scoring.
We did not perform significance testing; comparisons below are descriptive and intended to characterize performance level and variability in this setting.

\begin{table}[h]
\centering
\caption{\textbf{Aggregate LLM Backbone Comparison.} Overall mean accuracy, and per-report standard deviation (Std) on the Sugo dataset, benchmarked against the verified ground truth.}
\label{tab:result_table}
\small
\begin{tabular}{lcc}
\toprule
\textbf{Model Backbone} & \textbf{Mean Accuracy (\%)} & \textbf{Std (\%)} \\
\midrule

\textbf{gpt-oss:20b} & \textbf{86.02} & \textbf{6.87} \\
llama3-8b & 80.53 & 4.58 \\
mistral-7b & 78.89 & 4.68 \\
\rowcolor{Gray}
Clinical RA & 98.40 & 2.13 \\
\bottomrule
\end{tabular}
\vspace{-2mm}
\end{table}

Our quantitative results are summarized in Table~\ref{tab:result_table}. Using the same expert-sonographer annotated and double-checked reference labels (\emph{Verified Truth}), we scored and compared three locally-deployed LLMs and a Clinical Research Assistant (Clinical RA) performing manual abstraction. The Clinical RA achieved a mean accuracy of \textbf{98.40\%} with low variability (SD 2.13\%). Among the three LLMs, \texttt{gpt-oss:20b} achieved the highest mean accuracy in our dataset (\textbf{86.02\%}), while \texttt{llama3-8b} and \texttt{mistral-7b} achieved mean accuracies of 80.53\% and 78.89\%, respectively. Notably, while \texttt{gpt-oss:20b} achieved the highest mean accuracy among the LLMs, it also showed the largest per-report variance (SD 6.87\%), indicating less consistent performance across reports.

\begin{figure}[htbp]
  \centering
  \includegraphics[width=\columnwidth]{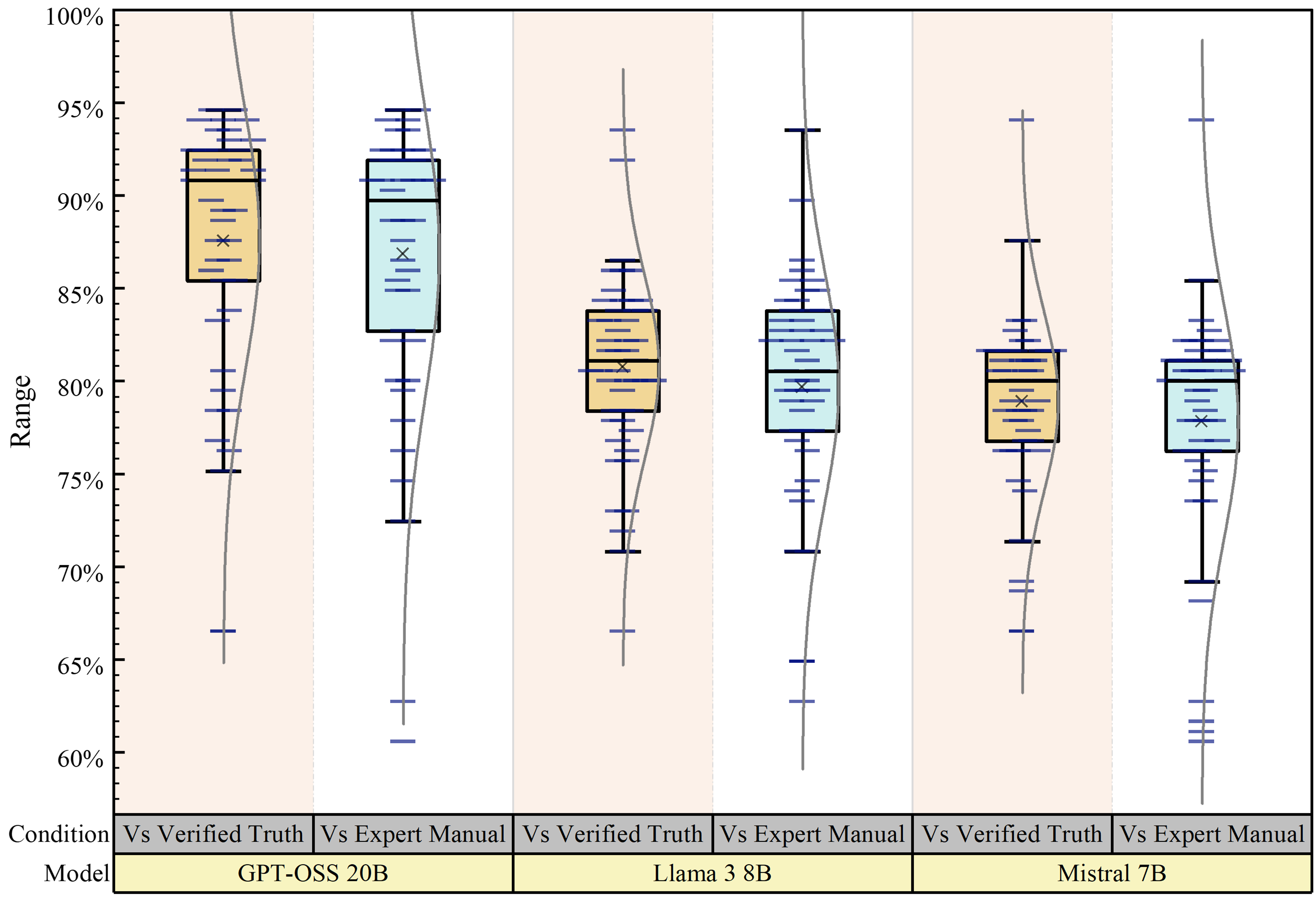} 
  \caption{Report-level accuracy distributions for three locally-deployed LLMs. \texttt{gpt-oss:20b} attains a higher median accuracy but exhibits larger variability, including occasional low-accuracy outliers.}
  \Description{Box-and-whisker plot of report-level accuracy for three models. GPT-OSS 20B shows a higher median but a wider interquartile range and more outliers than Llama 3 8B and Mistral 7B.}
  \label{fig:model_variance}
  \vspace{-4mm}
\end{figure}

Figure~\ref{fig:model_variance} further illustrates these distributional differences. The box-and-whisker plot shows that \texttt{gpt-oss:20b} attains a higher median accuracy but also a wider interquartile range (IQR), with a small number of outliers where accuracy drops markedly (e.g., below 65\%). Overall, this suggests that the larger model performs better on average in our dataset but is more sensitive to a subset of challenging reports, whereas smaller models exhibit a narrower (but lower) performance range.

\subsection{Error Analysis by Field Type}
\label{sec:Error Analysis}
\begin{figure}[h]
    \includegraphics[width=\columnwidth]{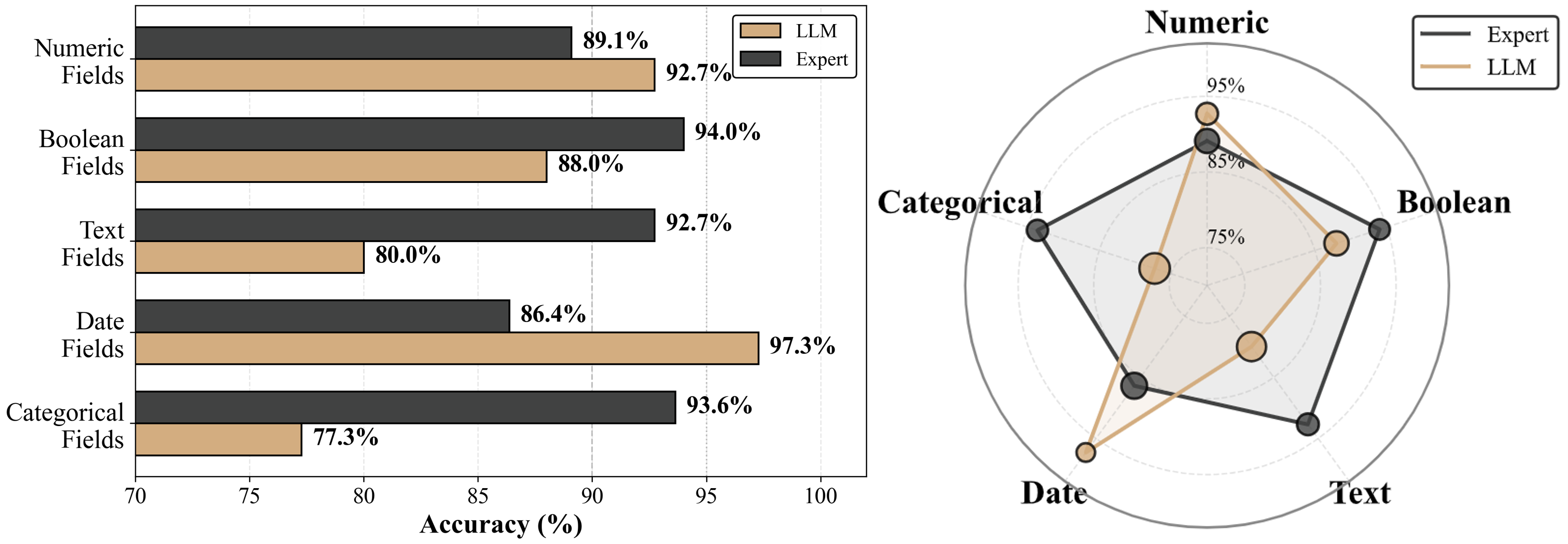}
    \caption{\textbf{Performance breakdown by field type.} In our setting, the LLM performs best on structured fields (Date, Numeric) and worse on semantically nuanced fields (Text, Categorical).}
    \label{fig:accuracy_by_field_type}
    \vspace{-4mm}
\end{figure}

To investigate the sources of divergence, we stratified performance by field data type (Date, Numeric, Categorical, Text), as shown in Figure~\ref{fig:accuracy_by_field_type}. Unless noted otherwise, the LLM results in this subsection focus on \texttt{gpt-oss:20b} as the best-performing local backbone in our study, to highlight its typical error distribution.

The analysis reveals complementary failure modes between the LLM and the human extractor. The LLM performs best on more structured, protocol-constrained fields, achieving its highest accuracy on \textbf{Date Fields (97.3\%)} and \textbf{Numeric Fields (92.7\%)}. Remaining errors in these categories are primarily omissions and schema-alignment failures (e.g., incomplete decomposition of multi-dimensional measurements). In contrast, errors are more concentrated in semantically sensitive Text and Categorical fields, where the model often fails through omissions or inconsistent terminology/ontology mapping \cite{Wang2018,info:doi/10.2196/12239,Agrawal2023}.

In contrast, the human extractor's errors were rarely clinical misinterpretations, but instead were predominantly data-entry protocol failures. A common error involved correctly reading a 3D nodule measurement from the report but failing to split it across three separate required database fields.

\subsection{Error Analysis on Clinically Important Fields}
\label{sec:clinically_important_fields}
To examine whether prompt engineering could improve performance on key items, we conducted a follow-up experiment using a critical-field prompt for \texttt{gpt-oss:20b}. This prompt explicitly identified the seven most clinically critical fields, with the goal of improving extraction consistency for these items. 
The critical-field prompt achieved marginally higher mean accuracy (88.8\%, SD = 6.23) compared to the generic prompt (87.0\%, SD = 6.44).
The critical-field prompt produced only a small change in mean accuracy on these fields, and the effect was not stable relative to report-level variability. We therefore treat this intervention as limited in utility for this schema-constrained task: emphasizing importance alone does not reliably change model behavior, and more practical safeguards are likely to come from auditable workflow mechanisms (e.g., rule-based validation, risk-prioritized review, and structured error logging) \cite{agrawal2022large,Holste2023}.

\section{Discussion}
\label{sec:discussion}

\subsection{Complementary Failure Modes Enable Task Allocation}

Our results reveal that LLMs and human abstractors fail on different field types, establishing a basis for differentiated task allocation. The LLM achieved near-human accuracy on Date (97.3\%) and Numeric (92.7\%) fields, where errors were predominantly mechanical (incomplete measurement decomposition, format mismatches) and detectable through rule-based validation. In contrast, the LLM struggled with Categorical and Text fields (77.3\% and 80.0\%), which require interpreting negation, mapping synonymous terms to controlled vocabularies, and inferring clinical intent. These semantic errors are syntactically well-formed, confirming that schema compliance alone cannot ensure extraction quality.

This complementarity suggests that uniform human oversight is neither necessary nor efficient. Fields with high LLM reliability and rule-detectable failures can be auto-validated with exception-based review, while semantically sensitive fields should be flagged for mandatory human verification. Implementing this workflow requires interfaces that link each extracted value to its source text, enabling reviewers to verify semantic correctness without re-reading entire reports.

\subsection{Extraction Variance as a Deployment Criterion}

Mean accuracy alone obscures operational risk. The 20B model achieved the highest average accuracy (86.02\%) but also the widest variance (SD = 6.87\%), with outliers below 65\%. Inspection of these cases revealed two patterns: reports with atypical formatting triggered cascading failures, and reports dense with negated or conditional findings led to accumulated errors across categorical fields. Smaller models showed lower variance but at a lower accuracy level, suggesting more consistent but conservative outputs.

High variance translates to unpredictable review burden. A reviewer calibrated for occasional errors may overlook reports where a third of extractions fail. This observation supports mechanisms such as confidence-based triage, routing structurally atypical or low-confidence reports to full review. More broadly, model selection should be framed as a variance-accuracy trade-off: in some operational contexts, a smaller model with predictable, recoverable errors may prove more practical than a larger model whose sporadic failures are harder to detect.

\section{Limitation}
Our evaluation is bounded by a modest sample (49 reports) from a single institution, which may not capture variation in reporting styles across sites. Data-sovereignty constraints restricted evaluation to locally deployable models (7B–20B parameters); cloud-hosted or domain-specialized medical backbones (e.g., MedGemma) may exhibit different error patterns and warrant future comparison. Finally, our critical-field prompting experiment yielded null results, suggesting that for schema-constrained tasks where errors are semantic rather than attentional, prompt emphasis alone is insufficient—motivating workflow-level safeguards rather than instruction tuning as the more practical mitigation.

\section{Conclusion}
\label{sec:conclusion}
We present a systematic evaluation of locally deployed LLMs for structured extraction from endometriosis transvaginal ultrasound reports, addressing settings where data-sovereignty requirements preclude cloud-based processing. Three findings inform the design of human-AI abstraction workflows. First, the 20B model achieved the highest field-level accuracy (86.02\%) while remaining feasible for on-premise deployment, demonstrating that capable local models can support clinical extraction tasks. Second, LLMs and human abstractors exhibit complementary error patterns: LLMs excel on structured fields while humans outperform on interpretive fields, suggesting a division of labour rather than full automation or uniform oversight. Third, targeted prompt engineering yielded no meaningful improvement, indicating that prompt refinement alone cannot resolve errors on interpretive fields and that workflow-level safeguards are necessary. These findings support a human-in-the-loop workflow in which local LLMs generate structured drafts at scale, automated validation flags mechanical errors, and targeted human review addresses interpretive fields requiring clinical judgement. Future work should validate these patterns on multi-site datasets and develop lightweight, auditable mechanisms for confidence-based triage.

\bibliographystyle{ACM-Reference-Format}
\bibliography{refs}


\begin{thebibliography}{18}


\ifx \showCODEN    \undefined \def \showCODEN     #1{\unskip}     \fi
\ifx \showISBNx    \undefined \def \showISBNx     #1{\unskip}     \fi
\ifx \showISBNxiii \undefined \def \showISBNxiii  #1{\unskip}     \fi
\ifx \showISSN     \undefined \def \showISSN      #1{\unskip}     \fi
\ifx \showLCCN     \undefined \def \showLCCN      #1{\unskip}     \fi
\ifx \shownote     \undefined \def \shownote      #1{#1}          \fi
\ifx \showarticletitle \undefined \def \showarticletitle #1{#1}   \fi
\ifx \showURL      \undefined \def \showURL       {\relax}        \fi
\providecommand\bibfield[2]{#2}
\providecommand\bibinfo[2]{#2}
\providecommand\natexlab[1]{#1}
\providecommand\showeprint[2][]{arXiv:#2}

\bibitem[Agrawal et~al\mbox{.}(2022)]%
        {agrawal2022large}
\bibfield{author}{\bibinfo{person}{Monica Agrawal}, \bibinfo{person}{Stefan Hegselmann}, \bibinfo{person}{Hunter Lang}, \bibinfo{person}{Yoon Kim}, {and} \bibinfo{person}{David Sontag}.} \bibinfo{year}{2022}\natexlab{}.
\newblock \showarticletitle{Large language models are few-shot clinical information extractors}.
\newblock \bibinfo{journal}{\emph{arXiv preprint}} (\bibinfo{year}{2022}).
\newblock
\urldef\tempurl%
\url{https://arxiv.org/abs/2205.12689}
\showURL{%
\tempurl}
\newblock
\shownote{arXiv:2205.12689}.


\bibitem[Beede et~al\mbox{.}(2020)]%
        {beede2020human}
\bibfield{author}{\bibinfo{person}{Emma Beede}, \bibinfo{person}{Elizabeth Baylor}, \bibinfo{person}{Fred Hersch}, \bibinfo{person}{Anna Iurchenko}, \bibinfo{person}{Lauren Wilcox}, \bibinfo{person}{Paisan Ruamviboonsuk}, {and} \bibinfo{person}{Laura~M. Vardoulakis}.} \bibinfo{year}{2020}\natexlab{}.
\newblock \showarticletitle{A Human-Centered Evaluation of a Deep Learning System Deployed in Clinics for the Detection of Diabetic Retinopathy}. In \bibinfo{booktitle}{\emph{Proceedings of the 2020 CHI Conference on Human Factors in Computing Systems}} (Honolulu, HI, USA) \emph{(\bibinfo{series}{CHI '20})}. \bibinfo{publisher}{Association for Computing Machinery}, \bibinfo{address}{New York, NY, USA}, \bibinfo{pages}{1–12}.
\newblock
\showISBNx{9781450367080}
\href{https://doi.org/10.1145/3313831.3376718}{doi:\nolinkurl{10.1145/3313831.3376718}}


\bibitem[Brown et~al\mbox{.}(2020)]%
        {Brown2020}
\bibfield{author}{\bibinfo{person}{Tom Brown}, \bibinfo{person}{Benjamin Mann}, \bibinfo{person}{Nick Ryder}, \bibinfo{person}{Melanie Subbiah}, \bibinfo{person}{Jared~D Kaplan}, \bibinfo{person}{Prafulla Dhariwal}, {et~al\mbox{.}}} \bibinfo{year}{2020}\natexlab{}.
\newblock \showarticletitle{Language Models are Few-Shot Learners}. In \bibinfo{booktitle}{\emph{Advances in Neural Information Processing Systems}}, \bibfield{editor}{\bibinfo{person}{H.~Larochelle}, \bibinfo{person}{M.~Ranzato}, \bibinfo{person}{R.~Hadsell}, \bibinfo{person}{M.~F. Balcan}, {and} \bibinfo{person}{H.~Lin}} (Eds.), Vol.~\bibinfo{volume}{33}. \bibinfo{publisher}{Curran Associates, Inc.}, \bibinfo{pages}{1877--1901}.
\newblock
\urldef\tempurl%
\url{https://proceedings.neurips.cc/paper_files/paper/2020/file/1457c0d6bfcb4967418bfb8ac142f64a-Paper.pdf}
\showURL{%
\tempurl}


\bibitem[Bu\c{c}inca et~al\mbox{.}(2021)]%
        {bucinca2021trust}
\bibfield{author}{\bibinfo{person}{Zana Bu\c{c}inca}, \bibinfo{person}{Maja~Barbara Malaya}, {and} \bibinfo{person}{Krzysztof~Z. Gajos}.} \bibinfo{year}{2021}\natexlab{}.
\newblock \showarticletitle{To Trust or to Think: Cognitive Forcing Functions Can Reduce Overreliance on AI in AI-assisted Decision-making}.
\newblock  \bibinfo{volume}{5}, \bibinfo{number}{CSCW1}, Article \bibinfo{articleno}{188} (\bibinfo{date}{April} \bibinfo{year}{2021}), \bibinfo{numpages}{21}~pages.
\newblock
\href{https://doi.org/10.1145/3449287}{doi:\nolinkurl{10.1145/3449287}}


\bibitem[Busch et~al\mbox{.}(2025)]%
        {Holste2023}
\bibfield{author}{\bibinfo{person}{Felix Busch}, \bibinfo{person}{Lena Hoffmann}, \bibinfo{person}{Daniel~Pinto Dos~Santos}, \bibinfo{person}{Marcus~R Makowski}, \bibinfo{person}{Luca Saba}, \bibinfo{person}{Philipp Prucker}, {et~al\mbox{.}}} \bibinfo{year}{2025}\natexlab{}.
\newblock \showarticletitle{Large language models for structured reporting in radiology: past, present, and future}.
\newblock \bibinfo{journal}{\emph{European Radiology}} \bibinfo{volume}{35}, \bibinfo{number}{5} (\bibinfo{year}{2025}), \bibinfo{pages}{2589--2602}.
\newblock


\bibitem[Bussone et~al\mbox{.}(2015)]%
        {bussone2015role}
\bibfield{author}{\bibinfo{person}{Adrian Bussone}, \bibinfo{person}{Simone Stumpf}, {and} \bibinfo{person}{Dympna O'Sullivan}.} \bibinfo{year}{2015}\natexlab{}.
\newblock \showarticletitle{The Role of Explanations on Trust and Reliance in Clinical Decision Support Systems}. In \bibinfo{booktitle}{\emph{2015 International Conference on Healthcare Informatics}}. \bibinfo{pages}{160--169}.
\newblock
\href{https://doi.org/10.1109/ICHI.2015.26}{doi:\nolinkurl{10.1109/ICHI.2015.26}}


\bibitem[Cai et~al\mbox{.}(2019)]%
        {cai2019hello}
\bibfield{author}{\bibinfo{person}{Carrie~J. Cai}, \bibinfo{person}{Samantha Winter}, \bibinfo{person}{David Steiner}, \bibinfo{person}{Lauren Wilcox}, {and} \bibinfo{person}{Michael Terry}.} \bibinfo{year}{2019}\natexlab{}.
\newblock \showarticletitle{"Hello AI": Uncovering the Onboarding Needs of Medical Practitioners for Human-AI Collaborative Decision-Making}.
\newblock \bibinfo{journal}{\emph{Proc. ACM Hum.-Comput. Interact.}} \bibinfo{volume}{3}, \bibinfo{number}{CSCW}, Article \bibinfo{articleno}{104} (\bibinfo{date}{Nov.} \bibinfo{year}{2019}), \bibinfo{numpages}{24}~pages.
\newblock
\href{https://doi.org/10.1145/3359206}{doi:\nolinkurl{10.1145/3359206}}


\bibitem[Castro et~al\mbox{.}(2017)]%
        {castro2017automated}
\bibfield{author}{\bibinfo{person}{Sergio~M Castro}, \bibinfo{person}{Eugene Tseytlin}, \bibinfo{person}{Olga Medvedeva}, \bibinfo{person}{Kevin Mitchell}, \bibinfo{person}{Shyam Visweswaran}, \bibinfo{person}{Tanja Bekhuis}, {and} \bibinfo{person}{Rebecca~S Jacobson}.} \bibinfo{year}{2017}\natexlab{}.
\newblock \showarticletitle{Automated annotation and classification of BI-RADS assessment from radiology reports}.
\newblock \bibinfo{journal}{\emph{Journal of Biomedical Informatics}}  \bibinfo{volume}{69} (\bibinfo{year}{2017}), \bibinfo{pages}{177--187}.
\newblock


\bibitem[Davis et~al\mbox{.}(2013)]%
        {davis2013automated}
\bibfield{author}{\bibinfo{person}{Mary~F Davis}, \bibinfo{person}{Subramaniam Sriram}, \bibinfo{person}{William~S Bush}, \bibinfo{person}{Joshua~C Denny}, {and} \bibinfo{person}{Jonathan~L Haines}.} \bibinfo{year}{2013}\natexlab{}.
\newblock \showarticletitle{Automated extraction of clinical traits of multiple sclerosis in electronic medical records}.
\newblock \bibinfo{journal}{\emph{Journal of the American Medical Informatics Association}} \bibinfo{volume}{20}, \bibinfo{number}{e2} (\bibinfo{year}{2013}), \bibinfo{pages}{e334--e340}.
\newblock


\bibitem[Fitzpatrick and Ellingsen(2013)]%
        {fitzpatrick2013review}
\bibfield{author}{\bibinfo{person}{Geraldine Fitzpatrick} {and} \bibinfo{person}{Gunnar Ellingsen}.} \bibinfo{year}{2013}\natexlab{}.
\newblock \showarticletitle{A Review of 25 Years of CSCW Research in Healthcare: Contributions, Challenges and Future Agendas}.
\newblock \bibinfo{journal}{\emph{Comput. Supported Coop. Work}} \bibinfo{volume}{22}, \bibinfo{number}{4–6} (\bibinfo{date}{Aug.} \bibinfo{year}{2013}), \bibinfo{pages}{609–665}.
\newblock
\showISSN{0925-9724}
\href{https://doi.org/10.1007/s10606-012-9168-0}{doi:\nolinkurl{10.1007/s10606-012-9168-0}}


\bibitem[Leyh-Bannurah et~al\mbox{.}(2018)]%
        {leyh2018deep}
\bibfield{author}{\bibinfo{person}{Sami-Ramzi Leyh-Bannurah}, \bibinfo{person}{Zhe Tian}, \bibinfo{person}{Pierre~I Karakiewicz}, \bibinfo{person}{Ulrich Wolffgang}, \bibinfo{person}{Guido Sauter}, \bibinfo{person}{Margit Fisch}, {et~al\mbox{.}}} \bibinfo{year}{2018}\natexlab{}.
\newblock \showarticletitle{Deep learning for natural language processing in urology: state-of-the-art automated extraction of detailed pathologic prostate cancer data from narratively written electronic health records}.
\newblock \bibinfo{journal}{\emph{JCO Clinical Cancer Informatics}}  \bibinfo{volume}{2} (\bibinfo{year}{2018}), \bibinfo{pages}{1--9}.
\newblock


\bibitem[Li et~al\mbox{.}(2023)]%
        {Li2023}
\bibfield{author}{\bibinfo{person}{Yunxiang Li}, \bibinfo{person}{Zihan Li}, \bibinfo{person}{Kai Zhang}, \bibinfo{person}{Ruilong Dan}, \bibinfo{person}{Steve Jiang}, {and} \bibinfo{person}{You Zhang}.} \bibinfo{year}{2023}\natexlab{}.
\newblock \showarticletitle{Chatdoctor: A medical chat model fine-tuned on a large language model meta-ai (llama) using medical domain knowledge}.
\newblock \bibinfo{journal}{\emph{Cureus}} \bibinfo{volume}{15}, \bibinfo{number}{6} (\bibinfo{year}{2023}).
\newblock


\bibitem[Savova et~al\mbox{.}(2017)]%
        {savova2017deepphe}
\bibfield{author}{\bibinfo{person}{Guergana~K Savova}, \bibinfo{person}{Eugene Tseytlin}, \bibinfo{person}{Sean Finan}, \bibinfo{person}{Melissa Castine}, \bibinfo{person}{Timothy Miller}, \bibinfo{person}{Olga Medvedeva}, {et~al\mbox{.}}} \bibinfo{year}{2017}\natexlab{}.
\newblock \showarticletitle{DeepPhe: a natural language processing system for extracting cancer phenotypes from clinical records}.
\newblock \bibinfo{journal}{\emph{Cancer Research}} \bibinfo{volume}{77}, \bibinfo{number}{21} (\bibinfo{year}{2017}), \bibinfo{pages}{e115--e118}.
\newblock


\bibitem[Sheikhalishahi et~al\mbox{.}(2019)]%
        {info:doi/10.2196/12239}
\bibfield{author}{\bibinfo{person}{Seyedmostafa Sheikhalishahi}, \bibinfo{person}{Riccardo Miotto}, \bibinfo{person}{Joel~T Dudley}, \bibinfo{person}{Alberto Lavelli}, \bibinfo{person}{Fabio Rinaldi}, {and} \bibinfo{person}{Venet Osmani}.} \bibinfo{year}{2019}\natexlab{}.
\newblock \showarticletitle{Natural Language Processing of Clinical Notes on Chronic Diseases: Systematic Review}.
\newblock \bibinfo{journal}{\emph{JMIR Medical Informatics}} \bibinfo{volume}{7}, \bibinfo{number}{2} (\bibinfo{date}{27 Apr} \bibinfo{year}{2019}), \bibinfo{pages}{e12239}.
\newblock
\showISSN{2291-9694}
\href{https://doi.org/10.2196/12239}{doi:\nolinkurl{10.2196/12239}}
\newblock
\shownote{PubMed: 31066697. Also available at: http://medinform.jmir.org/2019/2/e12239/}.


\bibitem[Singhal et~al\mbox{.}(2023)]%
        {Singhal2023}
\bibfield{author}{\bibinfo{person}{Karan Singhal}, \bibinfo{person}{Shekoofeh Azizi}, \bibinfo{person}{Tao Tu}, \bibinfo{person}{S~Sara Mahdavi}, \bibinfo{person}{Jason Wei}, \bibinfo{person}{Hyung~Won Chung}, {et~al\mbox{.}}} \bibinfo{year}{2023}\natexlab{}.
\newblock \showarticletitle{Large language models encode clinical knowledge}.
\newblock \bibinfo{journal}{\emph{Nature}} \bibinfo{volume}{620}, \bibinfo{number}{7972} (\bibinfo{year}{2023}), \bibinfo{pages}{172--180}.
\newblock


\bibitem[Wang et~al\mbox{.}(2018)]%
        {Wang2018}
\bibfield{author}{\bibinfo{person}{Yanshan Wang}, \bibinfo{person}{Liwei Wang}, \bibinfo{person}{Majid Rastegar-Mojarad}, \bibinfo{person}{Sungrim Moon}, \bibinfo{person}{Feichen Shen}, \bibinfo{person}{Naveed Afzal}, \bibinfo{person}{Sijia Liu}, \bibinfo{person}{Yuqun Zeng}, \bibinfo{person}{Saeed Mehrabi}, \bibinfo{person}{Sunghwan Sohn}, {and} \bibinfo{person}{Hongfang Liu}.} \bibinfo{year}{2018}\natexlab{}.
\newblock \showarticletitle{Clinical information extraction applications: A literature review}.
\newblock \bibinfo{journal}{\emph{Journal of Biomedical Informatics}}  \bibinfo{volume}{77} (\bibinfo{year}{2018}), \bibinfo{pages}{34--49}.
\newblock
\showISSN{1532-0464}
\href{https://doi.org/10.1016/j.jbi.2017.11.011}{doi:\nolinkurl{10.1016/j.jbi.2017.11.011}}


\bibitem[Wang et~al\mbox{.}(2023)]%
        {Agrawal2023}
\bibfield{author}{\bibinfo{person}{Yuqing Wang}, \bibinfo{person}{Yun Zhao}, {and} \bibinfo{person}{Linda Petzold}.} \bibinfo{year}{2023}\natexlab{}.
\newblock \showarticletitle{Are large language models ready for healthcare? a comparative study on clinical language understanding}. In \bibinfo{booktitle}{\emph{Machine Learning for Healthcare Conference}}. PMLR, \bibinfo{pages}{804--823}.
\newblock


\bibitem[Weiss and Langlotz(2008)]%
        {Langlotz2006}
\bibfield{author}{\bibinfo{person}{David~L. Weiss} {and} \bibinfo{person}{Curtis~P. Langlotz}.} \bibinfo{year}{2008}\natexlab{}.
\newblock \showarticletitle{Structured Reporting: Patient Care Enhancement or Productivity Nightmare?}
\newblock \bibinfo{journal}{\emph{Radiology}} \bibinfo{volume}{249}, \bibinfo{number}{3} (\bibinfo{year}{2008}), \bibinfo{pages}{739--747}.
\newblock
\href{https://doi.org/10.1148/radiol.2493080988}{doi:\nolinkurl{10.1148/radiol.2493080988}}
\newblock
\shownote{PMID: 19011178}.


\end{thebibliography}

\end{document}